\documentclass[aps,pra,twocolumn]{revtex4}
\usepackage{epsfig,dsfont,amssymb,amsmath,amsthm,amsfonts,amsbsy,mathrsfs}
\usepackage{graphicx}
\usepackage{bm}
\begin{document}

\title{Transverse multi-mode effects on the performance of photon-photon gates}

\author{Bing He$^1$, Andrew MacRae$^1$, Yang Han$^{1,2}$, A. I. Lvovsky$^1$,
and Christoph Simon$^1$}
\affiliation{$^1$ Institute for Quantum Information Science
and Department of Physics and Astronomy, University of
Calgary, Calgary T2N 1N4, Alberta, Canada\\
$^2$ College of Science, National University of Defense
Technology, Changsha 410073, China}

\begin{abstract}
The multi-mode character of quantum fields imposes
constraints on the implementation of high-fidelity quantum
gates between individual photons. So far this has only been
studied for the longitudinal degree of freedom. Here we
show that effects due to the transverse degrees of freedom
significantly affect quantum gate performance. We also discuss potential
solutions, in particular separating the two photons in the transverse
direction.
\end{abstract}

\maketitle

\section{introduction}
Photons are attractive as carriers of quantum information
because they propagate fast over long distances and
interact weakly with their environment. Their utility for
quantum information processing applications such as quantum
repeaters \cite{repeaters} or quantum computing
\cite{nielsen} would be further enhanced if it was possible
to efficiently implement two-qubit gates between individual
photons. Such two-qubit gates can be implemented
probabilistically using just linear optics and photon
detection \cite{klm}, but strong photon-photon interactions
would allow much more direct and deterministic
implementations. Several approaches to the implementation
of interaction-based photon-photon gates have been
proposed, including proposals based on Kerr non-linearities
in fibers or crystals \cite{chuang}, on electromagnetically
induced transparency (EIT) in atomic ensembles
\cite{schmidt}, and on the interaction of both photons
with an individual quantum system \cite{turchette}. See Refs. \cite{birnbaum,chen,mat,adams,Lo} for recent experimental progress.

In the simplest case, an ideal controlled-phase gate performs the transformations $|0\rangle |0\rangle \rightarrow |0\rangle |0\rangle, |0\rangle |1\rangle \rightarrow |0\rangle |1\rangle, |1\rangle |0\rangle \rightarrow |1\rangle |0\rangle, |1\rangle |1\rangle \rightarrow e^{i\phi}|1\rangle |1\rangle$, where $|0\rangle$ and $|1\rangle$ are zero- and one-photon states respectively. In the present context, the phase $\phi$ acquired by the state $|1\rangle |1\rangle$ is due to the interaction of the two input photons. For quantum information processing applications it is desirable to achieve $\phi=\pi$. In the first theoretical papers the main focus was on how to achieve interactions that are sufficiently strong to allow large phase shifts. The photonic pulses were typically idealized as single-mode. Later it was realized that the (longitudinal) multi-mode character of
the pulses imposes important constraints on the
implementation of high-fidelity quantum gates
\cite{kojima,friedler1,andre,masalas,friedler2,shapiro,gea,marzlin}.
The phase shifts due to the interaction depend on the
relative position of the two photons, and take different
values over the pulses because the photons have to be
described as extended wave packets rather than point
particles \cite{inhom}. As a consequence, an initial product state
of the two photons is mapped by the interaction onto an
output state that exhibits unwanted entanglement in the
photons' external degrees of freedom. For large phase
shifts this leads to low fidelities for the simplest
quantum gate proposals \cite{gea}.

It has been suggested that this difficulty can be overcome
by more sophisticated quantum gate designs where the two
photons pass through each other. This can be achieved by
trapping one \cite{friedler1,chen} or both \cite{andre} of
the two photons, by having a counter-propagation geometry
\cite{masalas,friedler2}, or by considering two photons
with different group velocities \cite{marzlin}.

Here we show that having the photons pass through each other is not
sufficient on its own, due to the presence of the
transverse degrees of freedom. In short, the
interaction-induced phases also depend on the relative
transverse position of the two photons, which leads to
fidelity limitations that cannot be mitigated by counter-propagation. In
the following we describe these limitations in detail. We
also discuss potential solutions, in particular separating
the two wave packets in the transverse direction, which is
possible for non-linearities based on long-range
interactions.

\section{Interacting pulse evolution}

The general picture for photon-photon gate is the interaction between
two fields $\hat{\Psi}_1({\bf x},t)$ and $\hat{\Psi}_2({\bf x},t)$ in three spatial dimensions.
The field operators at $t=0$ are defined as $\hat{\Psi}_i({\bf x})=\frac{1}{\sqrt{V}}\sum_{\bf k}\hat{a}_{i,{\bf k}}e^{i{\bf k}\cdot {\bf x}}$, where $V$ is the quantization volume. They might describe photons in a Kerr medium or polaritons in an EIT medium. The field operators satisfy the equal-time commutation relation $[\hat{\Psi}_i({\bf x},t),\hat{\Psi}^{\dagger}_j({\bf x'},t)]=\delta_{ij} \delta({\bf x-x'})$. This means that photon absorption is negligible (e.g. in the EIT case the spectra of the pulses are well inside the transparency window), so the evolution of the interacting fields can be regarded as
unitary. Making the standard slowly varying envelope
and paraxial approximations, we have the effective Hamiltonian \cite{g-c-08}
($\hbar\equiv 1$ is adopted hereafter)
\begin{equation}
\hat{K}=\sum_{j}\int d^3 x\hat{\Psi}_j^{\dagger}({\bf x},t)
\{ v\frac{1}{i}\nabla_{z}-\frac{v\nabla^2_{T}}{2k}\}\hat{\Psi}_j({\bf x},t)
\vspace{-0.2cm}
\end{equation}
to describe their free evolution. Here $v$ is the group
velocity in the positive or negative $z$ direction, and $k=\frac{2\pi}{\lambda}$ is
the carrier wave vector, $\nabla^2_{T}$ is the
transverse Laplace operator. The interaction for the two fields is \cite{fetter}
\begin{eqnarray}
\hat{V}&=&\sum_{i,j}\frac{1}{2}\int d^3x_1\int d^3x_2 \{\hat{\Psi}_i^{\dagger}({\bf x}_1,t)
\hat{\Psi}_j^{\dagger}({\bf x}_2,t)\Delta({\bf x}_1-{\bf x}_2)\nonumber\\
&&\hat{\Psi}_j({\bf x}_2,t)\hat{\Psi}_i({\bf x}_1,t)\},
\label{interaction}
\end{eqnarray}
where the terms of $i=j$ and $i\neq j$ in the sum correspond to self-phase and cross-phase modulation effect, respectively. Given the total Hamiltonian $\hat{H}=\hat{K}+\hat{V}$, the equation of motion, $i\partial_t \hat{\Psi}_i({\bf x},t)=[\hat{\Psi}_i({\bf x},t), \hat{H}]$,
for the counter-propagating fields read
\begin{eqnarray}
&&(\frac{\partial}{\partial t}+v \frac{\partial}{\partial
z_1}-iv\frac{\nabla^2_{T,1}}{2k})\hat{\Psi}_1({\bf
x}_1,t)=-i\hat{\alpha}({\bf x}_1,t)\hat{\Psi}_1({\bf
x}_1,t)\nonumber\\
&&(\frac{\partial}{\partial t}- v \frac{\partial}{\partial
z_2}-iv\frac{\nabla^2_{T,2}}{2k})\hat{\Psi}_2({\bf
x}_2,t)=-i\hat{\alpha}({\bf x}_2,t)\hat{\Psi}_2({\bf x}_2,t)\nonumber\\
\label{eqofmotion}
\end{eqnarray}
with the interaction potential
\begin{equation}
\hat{\alpha}({\bf x}_i,t)=\int d^3 x' \Delta({\bf x}_i-{\bf
x'}) (\hat{I}_1({\bf x'},t)+\hat{I}_2({\bf x'},t)),
\label{alpha}
\vspace{-0.2cm}
\end{equation}
where $\hat{I}_n({\bf x})=\hat{\Psi}^{\dagger}_n({\bf
x})\hat{\Psi}_n({\bf x})$.

A photon-photon gate is implemented by evolving an input bi-photon state $|\Phi\rangle=|1\rangle_1 |1\rangle_2=\int d^3 x_1 f_1({\bf x}_1) \hat{\Psi}^\dagger_1 ({\bf x}_1) \int d^3 x_2 f_2({\bf x}_2) \hat{\Psi}^\dagger_2 ({\bf x}_2) |0\rangle$, where $f_i({\bf x})= \langle 0|\hat{\Psi}_i({\bf x})|1\rangle$ are the pulse profiles, under
the unitary time evolution
\begin{eqnarray}
\hat{U}(t)=\mathbb{T}e^{-i\int_0^{t}dt'\hat{H}(t')}
=e^{-i\int_0^t dt' \hat{K}}
e^{-i\int_0^t dt'\hat{V}}e^{-i\hat{C}}
\label{evolution},
\end{eqnarray}
where $\mathbb{T}$ denotes the time-ordering operation, and the operator is factorized
using the Baker-Campbell-Hausdorff formula.
The first factor in Eq. (\ref{evolution}) describes the free evolution including
pulse propagation and pulse diffraction.
The second factor is from the interaction between pulses. The third factor contains all commutators between the exponents of the first two terms; they reflect the interplay between pulse motion and pulse interaction, which generally changes the pulse profiles.

The ideal output state under a gate operation would be $e^{i\phi} e^{-i\int_0^t dt' \hat{K}}|\Phi\rangle$, with $\phi$ being a homogeneous controlled phase, where we are taking into account the free evolution of the photons. The actual output state, however, will be $\hat{U}(t)|\Phi\rangle=\int d^3 x_1 \int d^3 x_2 \psi({\bf x}_1,{\bf
x}_2,t) \hat{\Psi}^\dagger_1 ({\bf x}_1) \hat{\Psi}^\dagger_2 ({\bf x}_2) |0\rangle $ \cite{out}, giving the two-particle wave function $\psi({\bf x}_1, {\bf x}_2,t)\equiv\langle 0|\hat{\Psi}_1({\bf x}_1,t)
\hat{\Psi}_2({\bf x}_2,t)|\Phi\rangle$, which is generally non-factorisable with respect to ${\bf x}_1$ and ${\bf x}_2$. The fidelity $F$ and the controlled phase $\phi$ of a gate operation are determined via the overlap between the actual output state and the freely evolved state $|\Phi_R\rangle=e^{-i\int_0^t dt' \hat{K}}|\Phi\rangle$:
\begin{eqnarray}
\sqrt{F}e^{i\phi}&=& \langle \Phi_R|\hat{U}(t)|\Phi\rangle=\langle \Phi|e^{-i\int_0^t dt'\hat{V}}e^{-i\hat{C}}|\Phi\rangle\nonumber\\
&=&\int d^3 x_1 d^3 x_2 \psi_0^*({\bf
x}_1,{\bf x}_2,t) \psi({\bf x}_1,{\bf x}_2,t),
\label{Fphi-delta}
\end{eqnarray}
where  $\psi_0({\bf x}_1,{\bf x}_2,t)$ is the corresponding two-particle function for the freely evolved state $|\Phi_R\rangle$.

The field equations (\ref{eqofmotion}) allow one to obtain the evolution of $\psi({\bf x}_1,{\bf x}_2,t)$ by multiplying $\hat{\Psi}_2({\bf x}_2,t)$ to the right of the first equation of (3) and $\hat{\Psi}_1({\bf x}_1,t)$ to the left of the second; then, the product with $\langle 0|$ and $|\Phi\rangle$ is taken on the addition of the equations, yielding the following linear equation for the two-particle function $\langle 0|\hat{\Psi}_1({\bf x}_1,t)\hat{\Psi}_2({\bf x}_2,t)|\Phi\rangle$:
\begin{eqnarray}
&&(\frac{\partial}{\partial t}+ v\frac{\partial}{\partial z_1})\langle 0|\hat{\Psi}_1({\bf x}_1,t)\hat{\Psi}_2({\bf x}_2,t)|\Phi\rangle\nonumber\\
&+&(\frac{\partial}{\partial t}- v\frac{\partial}{\partial z_2})\langle 0|\hat{\Psi}_1({\bf x}_1,t)\hat{\Psi}_2({\bf x}_2,t)|\Phi\rangle\nonumber\\
&-&(iv\frac{\nabla^2_{T,1}}{2k_0}+iv\frac{\nabla^2_{T,2}}{2k_0})\langle 0|\hat{\Psi}_1({\bf x}_1,t)\hat{\Psi}_2({\bf x}_2,t)|\Phi\rangle\nonumber\\
&=&-i\Delta({\bf x}_2-{\bf x}_1)\langle 0|\hat{\Psi}_1({\bf x}_1,t)\hat{\Psi}_2({\bf x}_2,t)|\Phi\rangle.
\label{2p}
\end{eqnarray}
Here we have used $\Psi_i({\bf x})|0\rangle =0$, $\Psi_i({\bf x},t)\Psi_i^{\dagger}({\bf x'},t)|0\rangle =\frac{1}{V}\sum_{{\bf k}}e^{i{\bf k}\cdot({\bf x}-{\bf x}')}|0\rangle =\delta^{(3)}({\bf x}-{\bf x}')|0\rangle $.
With the center of mass coordinate, ${\bf X}=\frac{{\bf x}_1+{\bf x}_2}{2}$ and ${\bf x}={\bf x}_1-{\bf x}_2$, the derived equation for $\psi({\bf x}_1,{\bf x}_2,t)=R({\bf X},t)\xi({\bf x},t)$ can be reduced to
\begin{equation}
\{\frac{\partial}{\partial t}+2v\frac{\partial}{\partial z}+iv\frac{\nabla^2_{T,x}}{2k}\}\xi({\bf x},t)=-i\Delta({\bf x})\xi({\bf x},t),
\label{relative}
\vspace{-0cm}
\end{equation}
with $R({\bf X},t)$ being trivially evolved under the diffraction term $e^{-i\frac{vt}{k}\nabla^2_{T,X}}$.

The above linear equation greatly simplifies the determination of the evolution of interacting pulses. This simplification is possible because we are considering the interaction between two single photons (as opposed to multi-photon pulses). In the co-moving coordinate that eliminates the term $2v\partial_z\xi({\bf x},t)$ in
(\ref{relative}), the three factors in Eq. (\ref{evolution}) are translated into $e^{-i\frac{vt}{2k}\nabla^2_{T,x}}$, $e^{-i\varphi({\bf x},t)}$ and $e^{-i\hat{C}'}$, respectively, to evolve $\xi({\bf x},0)$. Now the third factor contains the exponentials of the commutators between $\frac{vt}{2k}\nabla^2_{T,x}$ and $\varphi({\bf x},t)$ as follows:
\begin{eqnarray}
e^{-i\hat{C}'}&=&\exp\{-\frac{1}{2}[\varphi, \frac{vt}{2k}\nabla^2_{T,x}]\}\nonumber\\
&\times &\exp\{\frac{i}{3}[\varphi,[\varphi, \frac{vt}{2k}\nabla^2_{T,x}]]
+\frac{i}{6}[\frac{vt}{2k}\nabla^2_{T,x},[\varphi,\frac{vt}{2k}\nabla^2_{T,x}]\}\nonumber\\
&\cdots &
\label{correction}
\end{eqnarray}
These commutators in the exponentials are of order $l/r$, where $l=vt$ is the medium length and $r=k\sigma^2$ is the Rayleigh length, with $\sigma$ the transverse size of the pulses at $t=0$.
It is not difficult to achieve $l/r \ll 1$ in practice, making the effects from
the third factor insignificant. For simplicity we will not include the first and third factor in the two-particle functions derived below, but the first order contribution from the third factor will be considered in our numerical calculations. The second factor $e^{-i\varphi({\bf x},t)}$, where $\varphi({\bf x},t)=\int_0^t dt' \Delta({\bf x}_T, z-2v(t-t'))$, is the main concern of the present paper. We will explain its effects with two examples.

\begin{figure}
\epsfig{file=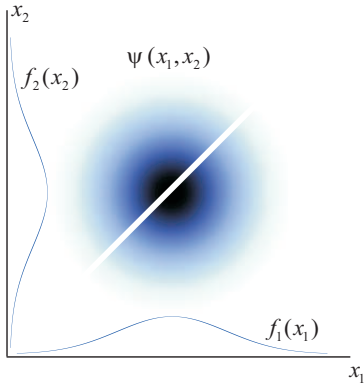,width=0.6\linewidth}
\caption{\vspace{-0.1cm} This qualitative plot shows the two-particle wave
function   as a function of the transverse coordinates $x_1$ and
$x_2$ of the two photons. The photons are distributed over a
multitude of transverse positions. For a contact interaction
the interaction induces a non-zero phase only if the two photons
happen to be at exactly the same position (the diagonal
line $x_1 = x_2$ in the figure), see Eq. (10). The probability of this
occurring is infinitesimal. As a consequence, the effective output
phase defined by Eq. (6) is always zero, independently
of the strength of the interaction.
}
\vspace{-0.35cm} \label{head-on}
\end{figure}

\section{Contact potential}

Our first example is a highly
local interaction described by a delta function potential,
$\Delta({\bf x}_1-{\bf x}_2)=V_0 \delta^{(3)}({\bf
x}_1-{\bf x}_2)$. This is a good model for non-linearities
that are due to the interaction of both photons with the
same atom in an atomic ensemble or crystal, or to short-range atomic
collisions \cite{chuang,schmidt,friedler1,masalas,andre}.
This interaction gives an output two-particle function
\begin{eqnarray}
&&\psi({\bf x}_1,{\bf x}_2,t)=f_1(z_1-vt,{\bf
x}_{T,1})f_2(z_2+vt,{\bf x}_{T,2})\nonumber\\
&& \times \exp\{i\frac{V_0}{2v} [H(z-2vt)-H(z)]\delta^{(2)}({\bf x}_T)\},
\label{output-delta}
\end{eqnarray}
where $H(z)$ is the Heaviside step function. From Eq. (\ref{output-delta}) one
sees that the interaction-induced phase is non-zero only if
the transverse coordinates of the two photons coincide,
i.e. on a subset of configuration space ${\bf x}_T=0$ that has measure
zero, see Fig. 1. As a consequence, one has $F=1$
and $\phi=0$. In the case of an ideal
three-dimensional delta function potential the effective output phase is exactly zero,
no matter how strong the interaction between the two photons.
This is closely related to the results of Refs.
\cite{kojima,shapiro} for the one-dimensional, but
co-propagating case. One finds essentially equivalent
results for any interaction whose range is much shorter
than the transverse size of the wave packets.
Note that this result is consistent with the non-zero conditional phase for photon-photon interactions obtained in Refs. \cite{lukin-imamoglu, petrosyan}, where
the evolution is non-unitary, as manifested by a different field operator commutator.

\begin{figure}
\epsfig{file=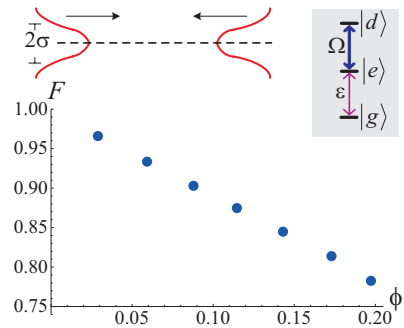,width=0.6\linewidth}
\caption{\vspace{-0.1cm} Schematic setup for a photon-photon gate working
with counter-propagating single photon pulses in a medium
under EIT conditions. The pulses interact with each other
through the dipole-dipole force between the Rydberg states
$|d_i\rangle$. The inset shows the relevant energy levels
of the atoms. The pulses collide head-on, but they have a
transverse extent $\sigma$. This leads to a dependence of
the interaction-induced phase for the two-particle wave
function on the relative transverse position, resulting in
a trade-off between the effective phase $\phi$ of the
two-photon operation and its fidelity $F$. Only very small
phases are compatible with high fidelities.}
\vspace{-0.35cm} \label{head-on}
\end{figure}

\section{Dipole-dipole interaction}

Related difficulties arise
also for more long-range interactions. This can be seen
from our second example, which is motivated by Refs.
\cite{friedler2,adams}. It concerns the interaction between
polaritons whose atomic component is in a highly excited
Rydberg state in an external electric field, cf. Fig.
\ref{head-on}. This induces a dipole-dipole interaction
between the polaritons,
\begin{equation}
\Delta({\bf x}_1-{\bf x}_2)=C(1-3\cos^2\vartheta)/|{\bf
x}_1-{\bf x}_2|^3,
\vspace{-0.1cm}
\end{equation}
where $C$ depends on the specific Rydberg states used and
$\vartheta$ is the angle between ${\bf x}_1-{\bf x}_2$ and
the external field (along which the electric dipoles of the
Rydberg states are aligned). This is an attractive system
because Rydberg states have large dipole moments, leading
to potentially very strong interactions between the
polaritons \cite{saffman,urban,gaetan}.

We consider the situation where the external field is
perpendicular to the direction of motion. We assume the initial pulse profiles to be $f_1({\bf x}_1)=\psi_0(x_1)\psi_0(y_1)\psi_0(z_1)$ and $f_2({\bf x}_2)=\psi_0(x_2)\psi_0(y_2)
\psi_0(z_2-l)$, where  $\psi_n(x)=[\frac{1}{\sigma\sqrt{\pi}2^n n!}]^{\frac{1}{2}}H_n(\frac{x}{\sigma})e^{-\frac{1}{2}(\frac{x}{\sigma})^2}$ with $H_n(\frac{x}{\sigma})$ being the Hermite polynomials.
The evolution according to Eq. (\ref{relative}) gives the output
two-particle wave function
\begin{eqnarray}
f_1(z_1-l,{\bf x}_{T,1})f_2(z_2+l,{\bf x}_{T,2})
e^{-i\varphi({\bf x}_{1}, {\bf x}_{2},l/v)}.
\end{eqnarray}
The interaction-induced phase in the above is given by
\begin{eqnarray}
\varphi(z,{\bf x}_T,l/v)&=&\frac{C}{2v}\frac{1}{{\bf x}_{T}^2}
\{\frac{z^3+2z{\bf x}_{T}^2}{(z^2+{\bf x}_T^2)^{\frac{3}{2}}}\nonumber\\
&-&\frac{(z-2l)^3+2(z-2l){\bf x}_{T}^2}{((z-2l)^2+{\bf x}_{T}^2)^{\frac{3}{2}}}\}.
 \label{phi-rydberg}
\end{eqnarray}

By Eq. (\ref{Fphi-delta}), the conditional phase $\phi$ and the fidelity $F$ in this case are determined as follows:
\begin{eqnarray}
&&\sqrt{F}e^{i\phi}= \int d^3x'_1 d^3x'_2 f^2_1({\bf x}'_1)f^2_2({\bf x}'_2)\exp\{-i\frac{C}{2v}\frac{1}{{\bf x'}_{T}^2}\nonumber\\
&& \times  [\frac{(z'+l)^3+2(z'+l){\bf x'}_{T}^2}{((z'+l)^2+{\bf x'}_T^2)^{\frac{3}{2}}}
-\frac{(z'-l)^3+2(z'-l){\bf x'}_{T}^2}{((z'-l)^2+{\bf x'}_{T}^2)^{\frac{3}{2}}}]\},\nonumber\\
\label{dipole-fidelity}
\end{eqnarray}
where ${\bf x}'_1={\bf x}_1-l\hat{e}_z$, ${\bf x}'_2={\bf x}_2$, and ${\bf x}'={\bf x}'_1-{\bf x}'_2$. In the calculation we have chosen $l=4\pi \sigma$ and $\sigma=10 \lambda$.
The results are shown in Fig. \ref{head-on}. Increasing the parameter $C/(2v\sigma^2)$, which indicates the interaction strength, increases the effective output phase
$\phi$, but diminishes the output fidelity $F$. As a consequence,
significant phase shifts are completely out
of reach if one wants to achieve high fidelities. In the numerical calculations, we have included the first order correction due to the third factor of Eq. (\ref{evolution}).
Such correction comes from the commutators $[\frac{l\nabla^2_{T}}{2k},\varphi]$ and $[[\frac{l\nabla^2_{T}}{2k},\varphi],\varphi]$, see Eq. (\ref{correction}). The commutators involving higher powers of $\varphi$ are shown to be vanishing. The exponentials of these commutators effect a modification of the intensity profile $f_i^2({\bf x}_i)$ and a modification of the phase profile $e^{-i\varphi(z,{\bf x}_T)}$, respectively.

The main cause for the trade-off between $\phi$ and $F$ is the dependence of the interaction-induced phase $\varphi(z,{\bf x}_T)$ on the transverse relative position ${\bf x}_T$. In fact, most of the behavior shown in Fig. \ref{head-on} can be understood by
setting the phase in the integrand of Eq. (\ref{dipole-fidelity}) as
$-\frac{C}{v} \frac{1}{{\bf x'}_{T}^2}$, which is quite accurate
for $l \gg \sigma$.
It gives rise to transverse mode mixing in the form
\begin{eqnarray}
&&e^{-i\varphi(z,{\bf x}_T)}\psi_0(x_1)\psi_0(y_1)\psi_0(x_2)\psi_0(y_2)\nonumber\\
&=&\sum_{m,n,l,k}C_{mnlk}\psi_m(x_1)\psi_n(y_1)\psi_l(x_2)\psi_k(y_2)\nonumber\\
&\neq &\{\sum_{m,n}c_{mn}\psi_m(x_1)\psi_n(y_1)\}\{\sum_{l,k}d_{lk} \psi_l(x_2)\psi_k(y_2)\},~~~~
\label{mixing}
\end{eqnarray}
leading to the deviation from the ideal output two-particle function $e^{i\phi}\psi_0(x_1)\psi_0(y_1)\psi_0(x_2)\psi_0(y_2)$.

The above analysis shows that the transverse mode mixing (or transverse mode entanglement) will develop even if the pulses are initially in a single transverse mode. This effect, which significantly affects the performance of photon-photon gates, is here described for the first time,
to the best of our knowledge. Our analysis also clarifies that the pulse diffraction in the transverse direction has no direct impact on the performance of a photon-photon gate. It influences gate performance through its interplay with the interaction between pulses, i.e., by the third factor in Eq. (\ref{evolution}). Compared with the effect of the transverse mode mixing shown in Eq. (\ref{mixing}), such diffraction-interaction interplay is insignificant in the regime considered here, where the medium length $l$ is much smaller than the Rayleigh length $r$. For example, for $l/r=0.2$ (as in our calculations) it induces corrections only at the few-percent level.

\begin{figure}
\epsfig{file=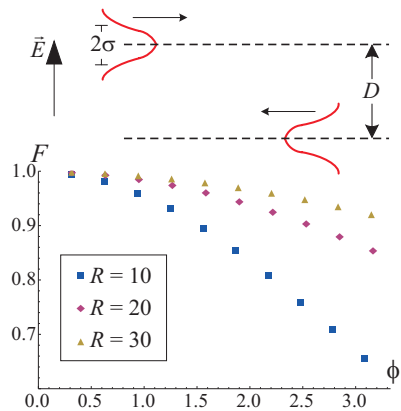,width=0.6\linewidth}
\caption{Introducing a transverse separation between the
two pulses greatly relaxes the trade-off between $F$ and
$\phi$, such that a phase of order $\pi$ becomes compatible
with high $F$. The dimensionless separation $R$ is defined
as $\frac{D}{\sigma}$. The price to pay is that the
interaction strength has to be increased significantly in
order to compensate for the transverse separation.}
\vspace{-0.35cm} \label{separated}
\end{figure}

\section{Potential solutions}

We have seen that the transverse multi-mode character of the
quantum fields, which leads to a transverse relative
position dependence of the interaction-induced phase
shifts, has very significant consequences for the fidelity
and phase achievable in photon-photon gates. We will now
discuss two potential solutions for this problem. The first
is applicable only to the case of long-range interactions.
It consists in separating the paths of the two photons by a
transverse distance $D$ that is much greater than the
transverse size $\sigma$ of the pulses, i.e., the initial profiles of the pulses will be,
for example, $f_1({\bf x}_1)=\psi_0(x_1)\psi_0(y_1)\psi_0(z_1)$ and $f_2({\bf x}_2)=\psi_0(x_2-D)\psi_0(y_2)
\psi_0(z_2-l)$. With increasing $D$ one will approach a situation where the transverse
degrees of freedom of the photons can effectively be treated as
point-like. As a consequence, the effect studied here will
diminish. This is shown in Fig. \ref{separated}. We have
again chosen a medium length $l=4\pi \sigma$.
Interaction-diffraction interplay effects are at or below
the $10^{-3}$ level in this case because they depend on the
gradients in the interaction-induced phase across the wave
packets, which decrease with increasing transverse
separation.

The price to pay for the improved fidelities shown in Fig.
\ref{separated} is that even for dipole-dipole interactions
the interaction-induced phase decreases quickly as the
transverse separation is increased. For example, achieving
a conditional phase shift $\phi=\pi$ with a fidelity
$F=0.9$ requires $R=\frac{D}{\sigma}=26$. Comparing to Ref.
\cite{friedler2}, this means that one could work with a
principal number of the Rydberg state $n \simeq 75$, a
transverse wave packet size $\sigma=7 \mu$m, and a group
velocity $v=4$ m/s. Achieving $\phi=\pi$ with a fidelity
$F=0.99$ requires $R=79$, which is possible provided that, for example, $n$ can be increased to 100, $v$ reduced to 1 m/s, and $\sigma$ reduced to 5 $\mu$m. These requirements are realistic with current technology, in particular Rydberg states with $n=79$ were already used in the
experiment of Ref. \cite{urban}.

The second potential solution, which is applicable both to
short-range and long-range interactions, consists in
imposing strong transverse confinement. If the confinement energy is much greater than the interaction energy, then excitations to higher-order transverse modes are largely suppressed.
All that the interaction can do in this case is multiply the lowest-order transverse mode by an almost uniform phase factor (a
non-uniform phase would imply non-negligible amplitudes in
higher-order transverse modes), thus allowing high-fidelity
quantum gates.  Sufficiently strong confinement could be achieved for example using hollow core photonic crystal fibers \cite{hollow} or optical nanofibers \cite{vetsch}.

\section{Summary}

The importance of the multi-mode character of quantum fields for the implementation of
photon-photon gates had been recognized in the past, but only
for the longitudinal degree of freedom. Here we showed
that the transverse degrees of freedom also play a significant role, imposing important constraints on the performance of potential quantum gates. For contact
interactions the effective phase is essentially always zero, no
matter how strong the interaction. For long-range
interactions the situation is more favorable, but there are still
significant trade-offs between the achievable phase and
fidelity. We discussed two potential solutions. One is to
have a significant transverse separation between the two
wave packets, which is possible only for long-range interactions. The price to pay is the need for an even stronger
interaction. The second potential solution is to impose
very strong transverse confinement, which may be
possible using hollow fibers or nanofibers. In any case it will
be essential for future implementations of photon-photon
gates to take transverse multi-mode effects into account.

\begin{acknowledgements}
We thank M. Afzelius, H. de Riedmatten, A. Rispe and B.
Sanders for useful discussions. This work was supported by
AI-TF, NSERC DG, CFI, AIF, {\it Quantum Works} and CIFAR.
\end{acknowledgements}

%


\begin{thebibliography}{99}
\bibitem{inhom} This undesirable effect was previously
discussed for the longitudinal degree of freedom under the
names of ``inhomogeneous phase shift'' and ``spectral
broadening''.
\bibitem{out}This actual output state $|\Phi_{out}\rangle=\hat{U}(t)|\Phi\rangle$ is from the definition of the two-particle function $\psi({\bf x}_1, {\bf x}_2,t)=\langle 0|\hat{U}^{\dagger}(t)\hat{\Psi}_1({\bf x}_1)
\hat{\Psi}_2({\bf x}_2)\hat{U}(t)|\Phi\rangle=\langle 0|\hat{\Psi}_1({\bf x}_1)
\hat{\Psi}_2({\bf x}_2)|\Phi_{out}\rangle$. $|\Phi_{out}\rangle$ is thus determined to the unique form by the relation $\langle 0|\hat{\Psi}_i({\bf x})\hat{\Psi}_i^{\dagger}({\bf x}')|0\rangle=\delta^{(3)}({\bf x}-{\bf x}')$.
\end{thebibliography}
\end{document}